\documentclass[twocolumn,showpacs,prd]{revtex4}
\usepackage{mathrsfs}
\usepackage{longtable,lscape}
\usepackage{txfonts}
\usepackage{amssymb}
\usepackage{indentfirst}
\usepackage{graphicx,,booktabs}
\usepackage{color}
\usepackage{amssymb}

\begin{document}
%\draft
%
%
\title{Charged bottomonium-like structures in the hidden-bottom dipion decays of $\Upsilon(11020)$}
%
%
%%% Authors list %%%
\author{Dian-Yong Chen$^{1,3}$}\email{chendy@impcas.ac.cn}
\author{Xiang Liu$^{1,2}$\footnote{Corresponding author}}
\email{xiangliu@lzu.edu.cn}
\author{Takayuki Matsuki$^{1,2,{4}}$}
\email{matsuki@tokyo-kasei.ac.jp}

\affiliation{$^1$Research Center for Hadron and CSR Physics,
Lanzhou University and Institute of Modern Physics of CAS, Lanzhou 730000, China\\
$^2$School of Physical Science and Technology, Lanzhou University, Lanzhou 730000,  China\\
$^3$Nuclear Theory Group, Institute of Modern Physics of CAS,
Lanzhou 730000, China\\
$^{4}$Tokyo Kasei University,
1-18-1 Kaga, Itabashi, Tokyo 173, JAPAN}

\date{\today}

\begin{abstract}

Under the Initial Single Pion Emission mechanism, we study the hidden-bottom dipion decays of $\Upsilon(11020)$, i.e.,
$\Upsilon(11020)\to \Upsilon(nS)\pi^+\pi^-$ $(n=1,2,3)$ and $\Upsilon(11020)\to h_b(mP)\pi^+\pi^-$ $(m=1,2)$. We predict
explicit sharp peak structures close to the $B\bar{B}^*$ and $B^*\bar{B}^*$ thresholds and their reflections in the $\Upsilon(1S)\pi^+$, $\Upsilon(2S)\pi^+$ and $h_b(1P)\pi^+$ invariant mass spectrum distributions. We suggest future experiment, i.e., Belle, BaBar, and forthcoming BelleII or Super-B, carry out the search for these novel
phenomena, which can provide important test to the Initial Single Emission mechanism existing in higher bottomonia.

\end{abstract}

%\medskip
%\preprint{}
\maketitle

Recently we have
proposed a new decay mechanism, Initial Single Pion
Emission (ISPE) \cite{Chen2011pv}, existing in the hidden-bottom
dipion decays of $\Upsilon(5S)$. By the ISPE mechanism, we
have succeeded in producing
two charged bottomonium-like structures in the $\Upsilon(nS)\pi^\pm$
$(n=1,2,3)$ and $h_b(mP)\pi^\pm$ $(m=1,2)$ invariant mass spectrum
distributions \cite{Chen2011pv}, which are above
the $B\bar{B}^*$ and $B^*\bar{B}$ thresholds, respectively. What is most important is
that these peak
structures appear exactly at the energies corresponding
to two charged
$Z_b(10610)$ and $Z_b(10650)$ newly observed by the Belle Collaboration
\cite{Collaboration2011gj}. In our previous work \cite{Chen:2011zv}, we introduced the intermediate $Z_b(10610)$ and $Z_b(10650)$ contribution to the $\Upsilon(5S)$ hidden-bottom dipion decay, where we have solved the puzzling on the $\cos\theta$ distribution of $\Upsilon(5S)\to \Upsilon(2S)\pi^+\pi^-$ given by Belle \cite{Abe:2007tk}. To some extent, the work presented in  Ref. \cite{Chen2011pv} adopted different assumptions compared with Ref. \cite{Chen:2011zv}.

Following Ref. \cite{Chen2011pv}, we
have applied the ISPE mechanism to
the hidden-charm dipion
decays of %higher charmonia
$\psi(4040)$, $\psi(4160)$, $\psi(4415)$,
and %charmonium-like state
$Y(4260)$, %. Further we
and two
charged charmonium-like structures are predicted
close to the $D\bar{D}^*$ and
$D^*\bar{D}^*$ thresholds \cite{Chen2011xk}. The predicted line
shape of $d\Gamma/d{m_{h_c(1P)\pi}}$ of $\psi(4160)\to
h_c(1P)\pi^+\pi^-$ can
explain
the CLEO-c measurement of the
$h_c(1P)\pi^\pm$ invariant mass distribution from $e^+e^-\to
h_c(1P)\pi^+\pi^-$ at $E_{CM}=4170$ MeV \cite{2011uqa}.

\begin{center}
\begin{figure}[htb]
\scalebox{1.05}{\includegraphics{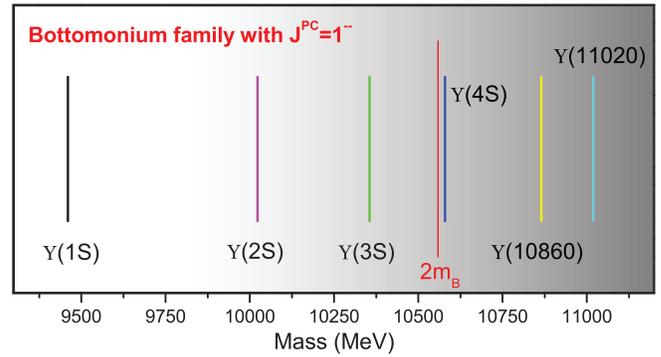}} \caption{(Color
online.) The mass spectrum of $\Upsilon$ family. Here, we also list
the threshold of $B\bar{B}$ and make a comparison with $\Upsilon$
states. \label{SP}}
\end{figure}
\end{center}

If the ISPE mechanism is a
universal one in the hidden-bottom
dipion decays of heavy quarkonia, we can naturally
apply
this to study the decay behaviors of other higher bottomonia,
and predict some novel phenomena. In Fig. \ref{SP}, we present the
mass spectra of bottomonia with $J^{PC}=1^{--}$
\cite{Nakamura2010zzi}, which are usually named as the $\Upsilon$
family. The comparison of $\Upsilon$ states with the $B\bar{B}$
threshold indicates that there are only three bottomonia just above
the $B\bar{B}$ threshold. Other than $\Upsilon(10860)$ studied in Ref.
\cite{Chen2011pv}, $\Upsilon(11020)$ can serve a good testing ground to
study the hidden-bottom dipion decay involving the ISPE mechanism,
where $\Upsilon(11020)$ was first observed by the CUSB Collaboration
\cite{Lovelock1985nb} and the CLEO Collaboration
\cite{Besson1984bd}, and measured again by BaBar recently
\cite{2008hx}.

\begin{figure}[htb]
\centering
\begin{tabular}{cc}
\scalebox{0.7}{\includegraphics{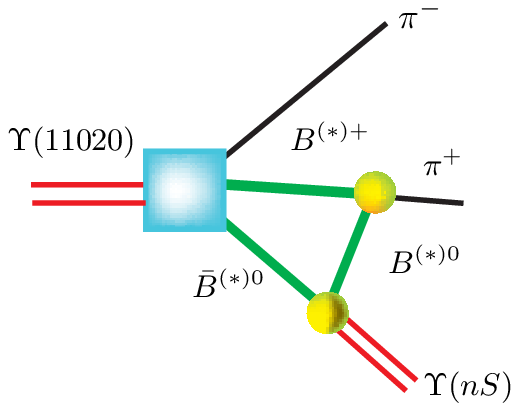}}&
\scalebox{0.7}{\includegraphics{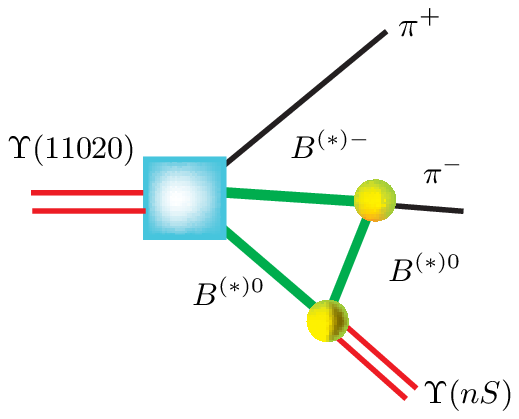}}\\
(a)&(b)
\end{tabular}
\caption{The schematic hadron-level diagrams relevant to
hidden-bottom dipion decay of $\Upsilon(11020)$ due to the ISPE
mechanism. By replacing $\Upsilon(nS)$ with $h_b(mP)$ in the final state, we obtain the
schematic diagrams for $\Upsilon(11020)\to h_b(mP)\pi^+\pi^-$.
\label{decay}}
\end{figure}

Because the mass of $\Upsilon(11020)$ is above the sum of the masses
of the emitted $\pi$ and the intermediate $B^{(*)}+\bar{B}^{(*)}$,
the pion initially emitted by $\Upsilon(11020)$ plays a crucial role
to make $B^{(*)}$ and $\bar{B}^{(*)}$ have low momenta, which can
easily interact with each other to transit into final states. This picture
is named as the ISPE mechanism \cite{Chen2011pv}. In Fig.
\ref{decay}, we give the schematic diagrams for %to describe
the hidden-bottom dipion decay of $\Upsilon(11020)$.

In this letter, the hidden-bottom dipion decays of $\Upsilon(11020)$
include
\begin{eqnarray}
\Upsilon(11020)\Rightarrow
\pi^\pm+\left\{\begin{array}{l}\{B\bar{B}\}^\mp\\\{B\bar{B}^*+B^*\bar{B}\}^\mp\\\{B^*\bar{B}^*\}^\mp\end{array}\right.
\Rightarrow \left\{\begin{array}{l}\Upsilon(1S)\pi^+\pi^-\\\Upsilon(2S)\pi^+\pi^-\\\Upsilon(3S)\pi^+\pi^-\\
h_b(1P)\pi^+\pi^-\\h_b(2P)\pi^+\pi^-\end{array}\right. ,
\end{eqnarray}
where $B\bar{B}$, $B\bar{B}^*+B^*\bar{B}$, and $B^*\bar{B}^*$ are
the intermediate states contributing to the triangle loops. The
superscript $\pm$ or $\mp$ denotes the charges of pion or
$B^{(*)}\bar{B}^{(*)}$ pair.
Diagrams (a) and (b) shown in Fig. \ref{decay} can be transformed into each other if considering
particle and antiparticle conjugation $B^{(*)}\rightleftharpoons
\bar{B}^{(*)}$ and $\pi^+\rightleftharpoons\pi^-$. By
interchanging $B^{(*)+}\rightleftharpoons B^{(*)0}$,
$B^{(*)-}\rightleftharpoons \bar{B}^{(*)0}$, and
$\pi^+\rightleftharpoons \pi^-$, one can deduce other diagrams
relevant to the hidden-bottom dipion decay of $\Upsilon(11020)$. We
find that there exist 4, 12, and 8 diagrams for $\Upsilon(11020)\to
\Upsilon(nS)\pi^+\pi^-$ decay via intermediate $B\bar{B}$,
$B\bar{B}^*+B^*\bar{B}$, and $B^*\bar{B}^*$, respectively, and 4, 8, and 8
diagrams for $\Upsilon(11020)\to h_b(mP)\pi^+\pi^-$ via intermediate
$B\bar{B}$, $B\bar{B}^*+B^*\bar{B}$, and $B^*\bar{B}^*$, respectively
(see Ref. \cite{Chen2011xk} for more details).

Because we use hadron-level description to the hidden-bottom dipion
decays of $\Upsilon(11020)$, the effective Lagrangian approach
is an appropriate way to describe the decay amplitudes relevant to this process.
The effective interaction Lagrangians
involved in our calculation are given by, \cite{Kaymakcalan:1983qq,Oh2000qr,Casalbuoni1996pg,Colangelo2002mj}
%%%%
%%%%
%%%%
%%%%
%%%%
\begin{eqnarray}
%%
%%Upsilon B(*) B(*) \pi
&&\mathcal{L}_{\Upsilon(11020) B^{(*)} B^{(*)} \pi} = \nonumber\\
&& -ig_{\Upsilon^\prime BB \pi} \varepsilon^{\mu \nu \alpha \beta}
\Upsilon_{\mu}^\prime \partial_{\nu} B \partial_{\alpha} \pi
\partial_{\beta} \bar{B} + g_{\Upsilon^\prime B^\ast B \pi} {\Upsilon^\prime}^{\mu}
(B \pi \bar{B}^\ast_{\mu} + B^\ast_{\mu} \pi \bar{B}) \nonumber\\
&&-ig_{\Upsilon^\prime B^\ast B^\ast \pi} \varepsilon^{\mu \nu \alpha
\beta} \Upsilon_{\mu}^\prime B^\ast_{\nu} \partial_{\alpha} \pi
\bar{B}^\ast_\beta -ih_{\Upsilon^\prime B^\ast B^\ast \pi} \varepsilon^{\mu
\nu \alpha \beta} \partial_{\mu} \Upsilon_{\nu}^\prime B^\ast_{\alpha} \pi
\bar{B}^\ast_{\beta},\nonumber\\\label{h1}\\
%%
%%B*B(*) pi
&&\mathcal{L}_{B^\ast B^{(\ast)} \pi} = \nonumber \\
&&ig_{B^\ast B \pi} (B^\ast_{\mu} \partial^\mu \pi \bar{B}-B
\partial^\mu \pi \bar{B}^\ast_{\mu})-g_{B^\ast B^\ast \pi}
\varepsilon^{\mu \nu \alpha \beta} \partial_{\mu} B^\ast_{\nu} \pi
\partial_{\alpha}
\bar{B}^\ast_{\beta}, \nonumber\\\label{h2}\\
%%
%%Upsilon B(*)B(*)
&&\mathcal{L}_{\Upsilon(nS) B^{(*)} B^{(*)}} =\nonumber\\
&& ig_{\Upsilon BB} \Upsilon_{\mu} (\partial^\mu B \bar{B}- B
\partial^\mu \bar{B})-g_{\Upsilon B^\ast B} \varepsilon^{\mu \nu
\alpha \beta}
\partial_{\mu} \Upsilon_{\nu} (\partial_{\alpha} B^\ast_{\beta} \bar{B}
\nonumber\\&& + B \partial_{\alpha}
\bar{B}^\ast_{\beta})-ig_{\Upsilon B^\ast B^\ast} \big\{
\Upsilon^\mu (\partial_{\mu} B^{\ast \nu} \bar{B}^\ast_{\nu}
-B^{\ast \nu} \partial_{\mu}
\bar{B}^\ast_{\nu}) \nonumber\\
&&+ (\partial_{\mu} \Upsilon_{\nu} B^{\ast \nu} -\Upsilon_{\nu}
\partial_{\mu} B^{\ast \nu}) \bar{B}^{\ast \mu} +
B^{\ast \mu}(\Upsilon^\nu \partial_{\mu} \bar{B}^\ast_{\nu} -
\partial_{\mu} \Upsilon^\nu \bar{B}^\ast_{\nu})\big\}, \nonumber\\\label{h3}\\
%%
%%hb BB
&&\mathcal{L}_{h_b(mP) B^{(*)} B^{(*)}}=\nonumber\\ && g_{h_b B^\ast
B} h_b^\mu ( B \bar{B}^\ast_{\mu}+ B^\ast_\mu \bar{B})+ ig_{h_b
B^\ast B^\ast} \varepsilon^{\mu \nu \alpha \beta}
\partial_{\mu} h_{b \nu} B^\ast_{\alpha} \bar{B}^\ast_{\beta}\label{h4}
\end{eqnarray}
with $\pi=\vec{\tau}\cdot \vec{\pi}$. In addition, we take
%
%\textcolor[rgb]{1.00,0.00,0.00}{
${{B}^{(*)}}=\left(B^{(*)+},B^{(*)0}\right)$ and
${\bar B^{(*)T}}=\left(B^{(*)-},\bar{B}^{(*)0}\right)$, which correspond to the
%}
%
bottom meson isodoublets. These effective Lagrangians presented in
Eqs. (\ref{h1})-(\ref{h3}) can be
strictly derived, terms including epsilon tensor from Ref. \cite{Kaymakcalan:1983qq}
and other terms from Ref. \cite{Oh2000qr},
by extending the symmetry from $SU(4)$ to $SU(5)$. % (see Appendix A for more details).
Here, the four-point vertex
$\Upsilon(11020)B^{(*)}\bar{B}^{(*)}\pi$ is described by Eq.
(\ref{h1}). The first term in Eq. (\ref{h1}) is similar to the
effective Lagrangian describing $\omega\to [\pi+ \rho]_{L=1}$,
except for the replacement of $\rho$ by an anti-symmetric
combination of two fields corresponding to the $B$ and $\bar{B}$
mesons. Thus, $\Upsilon^\prime\pi B B$ can be simplified as a
$1^-\to [0^-+1^-]_{L=1}$ basic process. The second term in Eq.
(\ref{h1}) reflects $1^-\to 0^-+[0^-+1^-]_{L=0}$, while the third
term in Eq. (\ref{h1}) is a $1^-\to \{0^-+[1^-+1^-]_{L=1}\}_{L=1}$
process. In Eq. (\ref{h1}), the fourth term is same as the third
term, except for parity violation coupling to $B^*B^*\pi$ due to
the $V-A$ interaction. Eq. (\ref{h2}) corresponds to the couplings
of $B^{(*)}$ and $\bar{B}^{(*)}$ mesons with pion, which are
P-wave interactions, i.e., $B^{(*)}\to[B^{(*)}+\pi]_{L=1}$. Eqs.
(\ref{h3}) and (\ref{h4}) show the interaction of bottomonia
$\Upsilon(nS)$ and $h_b(mP)$ with $B^{(*)}$ and $\bar{B}^{(*)}$
mesons, respectively, where $\Upsilon(nS)\to
[B^{(*)}+\bar{B}^{(*)}]_{L=1}$ interactions are typical P-wave
couplings, while $h_b(mP)\to [B^{(*)}+\bar{B}^{(*)}]_{L=0}$ are
S-wave interactions. In the explanation presented above, the
subscripts $L=0$ and $L=1$ denote the interactions of the
subsystems in the brackets $[...]$ and $\{...\}$ being S-wave and
P-wave couplings, respectively. The coupling constants in Eq.
(\ref{h2}) satisfy the relation $g_{B^\ast B^\ast \pi} =
\frac{g_{B^\ast B \pi}}{\sqrt{m_B m_{B^\ast}}} =\frac{2
g}{f_\pi}$. By using the branching ratio of $D^\ast \to D \pi$
measured by CLEO-c \cite{Anastassov2001cw} and $f_{\pi}=132$ MeV,
one gets $g=0.59$ \cite{Isola2003fh}. In Eq.~(\ref{h3}), there
also exists the relation
$$g_{\Upsilon BB} = g_{\Upsilon B^\ast B^\ast} \frac{m_B}{m_{B^\ast}} =g_{\Upsilon
B^\ast B} m_{\Upsilon} \sqrt{\frac{m_B}{m_{B^\ast}}}
=\frac{m_{\Upsilon}}{f_{\Upsilon}},$$
where $f_{\Upsilon}$ and $m_{\Upsilon}$ denote the decay constant and the mass of $\Upsilon(nS)$, respectively. In addition, the coupling constants in
Eq.~(\ref{h4}) are determined as
\begin{eqnarray}
g_{h_b BB^\ast} &=& -2g_1 \sqrt{m_{h_b} m_B m_{B^\ast}} , \ \ g_{h_b
B^\ast B^\ast} =2 g_1 \frac{m_{B^\ast}}{\sqrt{m_{h_b}}},
\end{eqnarray}
with $g_1=-\sqrt{\frac{m_{\chi_{b0}}}{3}} \frac{1}{f_{\chi_{b0}}}$, where
$m_{\chi_{b0}}$ and $f_{\chi_{b0}}$ are the mass and the decay constant of $\chi_{b0}(1P)$,
respectively \cite{Colangelo2002mj}.

\begin{figure}[htb]
\centering
\begin{tabular}{cc}
\scalebox{0.6}{\includegraphics{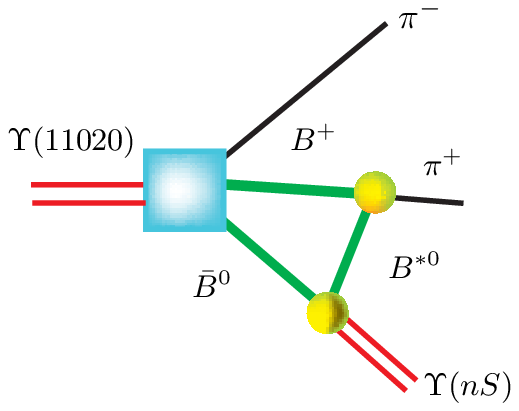}}&
\scalebox{0.6}{\includegraphics{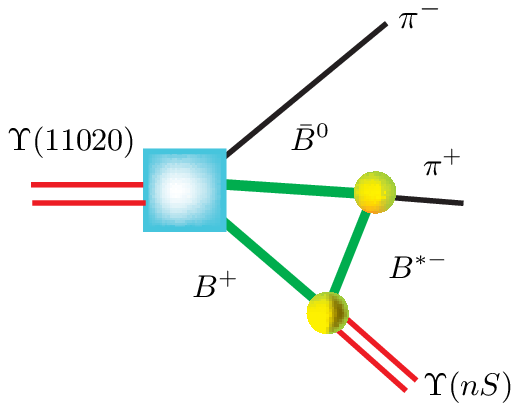}}\\
(a)&(b)\\
\scalebox{0.6}{\includegraphics{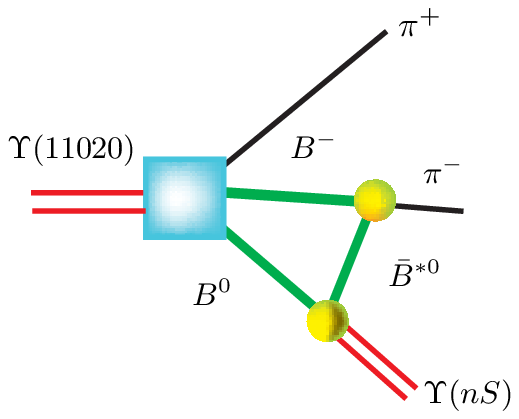}}&
\scalebox{0.6}{\includegraphics{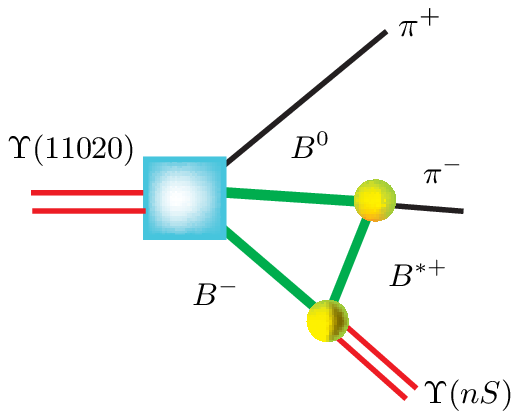}}\\
(c)&(d)
\end{tabular}
\caption{(Color Online) The hadron-level diagrams for
$\Upsilon(11020) \to \Upsilon(nS) \pi^+ \pi^-$ $(n=1,2,3)$ decays
with $B\bar{B}$ as the intermediate states. Replacing $\Upsilon(nS)$ with $h_b(mP)$, we get all diagrams for
$\Upsilon(11020)\to h_b(mP)\pi^+\pi^-$ decays. \label{S-channel} }
\end{figure}

Taking the process $\Upsilon(11020)\to \pi^{\pm} \{B\bar{B}\}^\mp\to \Upsilon(nS)\pi^+\pi^-,h_b(mP)\pi^+\pi^-$ as an example,
we illustrate how to deduce the corresponding decay amplitudes. In Fig. \ref{S-channel},
the ISPE mechanism gives the hadron-level diagrams depicting $\Upsilon(11020)\to \pi^{\pm} \{B\bar{B}\}^\mp\to \Upsilon(nS)\pi^+\pi^-$.
We can easily obtain the diagrams for
$\Upsilon(11020)\to \pi^{\pm} \{B\bar{B}\}^\mp\to h_b(mP)\pi^+\pi^-$
by replacing $\Upsilon(nS)$ with $h_b(mP)$.
%\begin{figure}[htb]
%\centering
%\begin{tabular}{cc}
%\scalebox{0.6}{\includegraphics{P-a.eps}}&
%\scalebox{0.6}{\includegraphics{P-b.eps}}\\
%(a)&(b)\\
%\scalebox{0.6}{\includegraphics{P-c.eps}}&
%\scalebox{0.6}{\includegraphics{P-d.eps}}\\
%(c)&(d)
%\end{tabular}
%\caption{(Color Online) The hadron-level diagrams for
%$\Upsilon(11020) \to h_b(mP) \pi^+ \pi^-, \{m=1,2\}$ decays with
%$B\bar{B}$ as the intermediate states. \label{P-channel}}
%\end{figure}

The decay amplitude of Fig. \ref{S-channel} (a) can be written as
\begin{eqnarray}
\mathcal{M}_a  &=&(i)^3 \int \frac{d^4 q}{(2\pi)^4}
[-ig_{\Upsilon^\prime BB \pi} \varepsilon_{\mu \nu \alpha \beta}
\epsilon_{\Upsilon^\prime}^\mu (ip_1^\rho) (ip_3^\alpha) (ip_2^\beta)]
\nonumber\\
&&\times [ig_{B^\ast B \pi} (-ip_4^\lambda)] [-g_{\Upsilon(nS)
B^\ast B} \varepsilon_{\delta \nu \theta \phi} (ip_5^\delta)
\epsilon_{\Upsilon(nS)}^\nu (-iq^\theta)]\nonumber\\
&&\times  \frac{1}{p_1^2-m_B^2} \frac{1}{p_2^2-m_B^2}
\frac{-g_{\lambda \phi} +q_{\lambda}
q_{\phi}/m_{B^\ast}^2}{q^2-m_{B^\ast}^2} \mathcal{F}^2(q^2),\label{f1}
\end{eqnarray}
or
\begin{eqnarray}
\mathcal{M}_a &=& (i)^3 \int \frac{d^4 q}{(2\pi)^4}
[-ig_{\Upsilon^\prime BB \pi} \varepsilon_{\mu \rho \alpha \beta}
\epsilon_{\Upsilon^\prime}^\mu (ip_1^\rho) (ip_3^\alpha) (ip_2^\beta)]
\nonumber\\
&&\times [ig_{B^\ast B \pi} (-ip_{4 \lambda})] [-g_{h_b B^\ast B}
\epsilon_{h_b \nu}] \frac{1}{p_1^2-m_B^2}\frac{1}{p_2^2-m_B^2}\nonumber\\
&&\times
\frac{-g^{\lambda \nu} +q^\lambda q^\nu/m_{B^\ast}^2}{
q^2-m_{B^\ast}^2} \mathcal{F}^2(q^2),\label{f2}
\end{eqnarray}
which corresponds to the process $$\Upsilon(11020)\to \pi^{-} B^+(p_1)\bar{B}^0(p_2)\to \Upsilon(nS)(p_3)\pi^{+}(p_4)\pi^-(p_5),$$
or
$$\Upsilon(11020)\to \pi^{-} B^+(p_1)\bar{B}^0(p_2)\to h_b(mP)(p_3)\pi^{+}(p_4)\pi^-(p_5),$$
where $q$ denotes the momentum of the exchanged meson $B^{*0}$ in transition $B^+\bar{B}^0\to \Upsilon(nS)\pi^{+}\pi^-$.
We take $\mathcal{F}(q^2)=(\Lambda^2-m_{B^*}^2)/(q^2-m_{B^{*}}^2)$, which denotes the monopole form factor
to describe the vertex structure appearing in the $B\bar{B}\to \Upsilon(nS)\pi^+$ transition.  In addition, such form factor also plays an important role to compensate the off-shell effect of the exchanged $B^{(*)}$ mesons. In general, the cutoff $\Lambda$
can be parameterized as $\Lambda=\xi\Lambda_{QCD}+m_{B^*}$ with $\Lambda_{QCD}=220$ MeV.
The decay amplitude of Fig. \ref{S-channel} (c) is given by Eq. (\ref{f1}) or Eq. (\ref{f2}) by interchanging $p_4$ and $p_5$ with each other,
i.e.,
$$\mathcal{M}_c=\mathcal{M}_a|^{p_4\to p_5}_{p_5\to p_4}.$$
Due to $SU(2)$ symmetry, we find other relations of the decay amplitudes
\begin{eqnarray}\mathcal{M}_b=\mathcal{M}_a,\quad \mathcal{M}_d=\mathcal{M}_c,\label{s}\end{eqnarray}
where decay amplitudes $\mathcal{M}_i$ $(i=a,b,c,d)$ correspond to Fig. \ref{S-channel} ($i$).  Eq. (\ref{s}) make the total decay amplitudes of $\Upsilon(11020)\to\Upsilon(nS)\pi^+\pi^-$
and $\Upsilon(11020)\to h_b(mP)\pi^+\pi^-$ be simplified as
\begin{eqnarray}
\mathcal{M}=2\left(\mathcal{M}_a+\mathcal{M}_c\right),
\end{eqnarray}
where the factor 2 reflects the $SU(2)$ symmetry just considered in this letter.
In Ref. \cite{Chen2011xk}, we have listed the detailed diagrams and formulation of the hidden-charm dipion decays of higher charmonia, where
we consider the intermediate $D\bar{D}^*+D^*\bar{D}$ and $D^*\bar{D}^*$ contributions to higher charmonium decays by the ISPE mechanism.
Thus, replacing \{initial higher charmonium$\to\Upsilon(11020)$\} and $\{D^{(*)}/\bar{D}^{(*)}\to \bar{B}^{(*)}/{B}^{(*)}\}$,
one can obtain the corresponding decay amplitudes for $\Upsilon(11020)$ hidden-bottom dipion decays via the intermediate $B\bar{B}^*+B^*\bar{B}$ and $B^*\bar{B}^*$ (see Ref. \cite{Chen2011xk} for more details).

With the above prescription, the differential decay width for $\Upsilon(11020)$ decay into $\Upsilon(nS) \pi^+ \pi^-$ reads as
\begin{eqnarray}
d\Gamma = \frac{1}{(2 \pi)^3} \frac{1}{32
m_{\Upsilon(11020)}^3} \overline{|\mathcal{M}|^2}
dm_{\Upsilon(nS) \pi^+}^2 dm_{\pi^+\pi^-}^2
\end{eqnarray}
with $m_{\Upsilon(nS) \pi^+}^2 = (p_4 + p_5)^2$ and $m_{\pi^+\pi^-}^2
=(p_3 +p_4)^2$, where the overline indicates the
average over the polarizations of the $\Upsilon(11020)$ in the
initial state and the sum over the polarization of $\Upsilon(nS)$ in
the final state. Replacing $m_{\Upsilon(nS)\pi^+}$ with $m_{h_b(mP)\pi^+}$, we obtain
the differential decay width for $\Upsilon(11020)
\to h_b(mP) \pi^+ \pi^-$. The values of the meson masses involved in the hidden-bottom dipion decays of
$\Upsilon(11020)$ are listed in Table. \ref{parameter}.
\begin{table}\caption{A summary of mass adopted in this letter. \label{parameter}}
\begin{tabular}{cccccccccccccc}
 \toprule[1pt]
\multicolumn{6}{c}{Mass (MeV) \cite{Nakamura2010zzi}} \\
 \midrule[1pt]
$\Upsilon(11020)$&$\Upsilon(3S)$&$\Upsilon(2S)$&$\Upsilon(1S)$&$h_b(1P)$ \cite{Adachi2011ji}&$h_b(2P)$
\cite{Adachi2011ji}\\
11019&10355&10023&9460& 9898 & 10259
\\
$B$&$B^*$&$\pi$&\\
5279&5325&140&\\%\midrule[1pt]
%\multicolumn{6}{c}{Coupling constant } \\
 \bottomrule[1pt]
\end{tabular}
\end{table}

\begin{figure*}[htbp]
\centering \scalebox{1.16}{\includegraphics{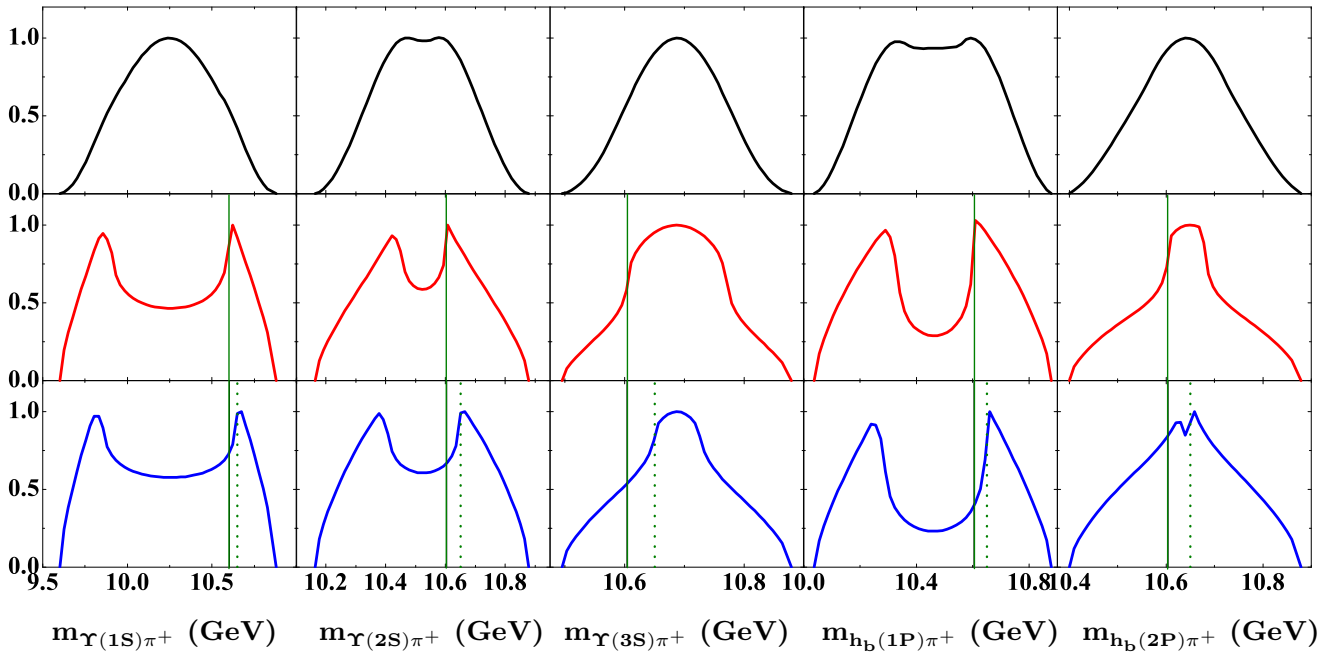}}
\caption{(Color online.) The distribution of $d\Gamma(\Upsilon(11020)\to\Upsilon(nS)
\pi^+\pi^+)/dm_{\Upsilon(nS)\pi^+}$ and $d\Gamma(\Upsilon(11020)\to h_b(mP)
\pi^+\pi^+)/dm_{h_b(mP)\pi^+}$ dependent on the $\Upsilon(nS)
\pi^+$ and $h_b(mP) \pi^+$ invariant mass spectra. Here, the results presented in the first, the second, and the third rows
are from the intermediate $B \bar{B}$, $B \bar{B^\ast} +B^\ast\bar{B}$, and $B^\ast \bar{B}^\ast$ contributions, respectively. The vertical solid and dashed
lines denote the threshold of $B\bar{B}^\ast+ B^\ast\bar{B}$ and
$B^\ast \bar{B}^\ast$, respectively. \label{diagram}}
\end{figure*}

In Fig. \ref{diagram}, we show the dependence of $d\Gamma(\Upsilon(11020)\to\Upsilon(nS)
\pi^+\pi^-)/dm_{\Upsilon(nS)\pi^+}$ and $d\Gamma(\Upsilon(11020)\to h_b(mP)
\pi^+\pi^-)/dm_{h_b(mP)\pi^+}$ on the $\Upsilon(nS)\pi^+$ and $h_b(mP) \pi^+$ invariant mass spectra, respectively,
where we take $\xi=1$. We need to specify that these numerical results listed in Fig. \ref{diagram} are weakly dependent on the values of the parameter $\xi$, which is consistent with that found in Ref. \cite{Chen2011pv}. Because we only focus on the line shapes of $d\Gamma(\Upsilon(11020)\to\Upsilon(nS)
\pi^+\pi^-)/dm_{\Upsilon(nS)\pi^+}$ and $d\Gamma(\Upsilon(11020)\to h_b(mP)
\pi^+\pi^-)/dm_{h_b(mP)\pi^+}$, the maxima of the line shapes in Fig. \ref{diagram} are normalized to be 1.

Our theoretical calculation indicates (1) there exist explicit sharp peaks close to the $B\bar{B}^*$ and $B^*\bar{B}^*$ thresholds in
the $\Upsilon(1S)\pi^+$, $\Upsilon(2S)\pi^+$, and $h_b(1P)\pi^+$ invariant mass spectrum distributions. In addition, we also find the reflections of these sharp peaks on the lower side of the invariant mass; (2) the broad structures close to the $B\bar{B}^*$ and $B^*\bar{B}^*$ thresholds appear in the $\Upsilon(3S)\pi^+$ invariant mass spectrum distribution, which are due to the overlapping peaks of two corresponding reflections; (3) in the $h_b(2P)\pi^+$ invariant mass spectrum, we also find a structure around $B\bar{B}^*$ threshold, which is narrower than that appearing in the $\Upsilon(3S)\pi^+$ invariant mass spectrum. In addition, we also find a small peak around $B^*\bar{B}^*$ and its reflection, which are close to each other; (4) the intermediate $B\bar{B}$ contribution to the hidden-bottom dipion decays of $\Upsilon(11020)$ does not give the phenomena similar to those from
the intermediate $B\bar{B}^*$ and $B^*\bar{B}^*$ states contributing to $\Upsilon(11020)$ decays just described above. They just give broad background-like line shape.

When comparing the results shown in Fig. \ref{diagram} with those of Figs. 3-4 in Ref. \cite{Chen2011pv}, we notice the differences between these results, which are mainly due to the change of mass of the initial state, i.e., the mass of $\Upsilon(11020)$ is different from that of $\Upsilon(5S)$ (10860 MeV).

The comparison of line shapes in Fig. \ref{diagram} indicates that the line shapes are dependent on the the definite hidden-bottom dipion decays of $\Upsilon(11020)$. Although we have obtained the peak structures just shown in Fig. \ref{diagram}, these are not typical Breit-Wigner type distributions when subtracting the contribution from the reflection. Thus, use of mass and width to specify these peak structures is not appropriate and realistic. In this work, we would like to emphasize that the sharp peak structures appear in the corresponding invariant mass spectrum, which can be tested in future experiment.

In conclusion, in this letter we study the hidden-bottom dipion decays of $\Upsilon(11020)$ by the ISPE mechanism, which has been first proposed in Ref. \cite{Chen2011pv} to study two charged $Z_b$ structures observed by Belle \cite{Collaboration2011gj} and has also been applied to investigate the hidden-charm dipion decays of higher charmonia \cite{Chen2011xk}. Furthermore, we predict the charged bottom-like structures close to the $B\bar{B}^*$ and $B^*\bar{B}^{*}$ thresholds, which exist in the $\Upsilon(nS)\pi^\pm$ and $h_b(mP)\pi^\pm$ invariant mass spectra of the hidden-bottom dipion decays of $\Upsilon(11020)$.
Just indicated in this letter, these charged peak structures are predicted due to the ISPE mechanism, a peculiar effect involved in the decays of higher bottomonia and/or charmonia \cite{Chen2011pv,Chen2011xk}.
We must admit that there exists interference between background and the ISPE contributions, where in this work we do not include background contribution due to our ignorance to background contribution. To some extend, such interference effect could bring some uncertainty to our prediction.

At present, there only exist the experimental measurements of the mass and width for $\Upsilon(11020)$ \cite{Lovelock1985nb,Besson1984bd,2008hx} while the information of its strong decay behaviors is still absent \cite{Nakamura2010zzi}. To some extent, the prediction presented in this letter could stimulate experimentalists' interest in carrying out further study on $\Upsilon(11020)$, especially on its hidden-bottom dipion decays. Besides Belle and BaBar, the forthcoming Belle II experiment \cite{belleii} will provide a good platform to study the properties of $\Upsilon(11020)$. In addition, the SuperB Factory has been approved by the Italian government in the last year \cite{superb}, which is also suitable to study the bottomonium decays. Here, we suggest these experimental search for the predicted charged bottomonium-like structures around the $B\bar{B}^*$ and $B^*\bar{B}^{*}$ thresholds by the hidden-bottom dipion decays of $\Upsilon(11020)$, which will provide a crucial test on the ISPE mechanism \cite{Chen2011pv}.

\vfil
%\section*{Acknowledgment}
\begin{acknowledgments}
This project is supported by the National Natural Science
Foundation of China under Grants Nos. 11175073, No. 11005129, No.
11035006, No. 11047606, the Ministry of Education of China (FANEDD
under Grant No. 200924, DPFIHE under Grant No. 20090211120029,
NCET under Grant No. NCET-10-0442, the Fundamental Research Funds
for the Central Universities), and the West Doctoral Project of
Chinese Academy of Sciences.
One of the authors (TM) would like to sincerely thank Prof. Xiang Liu
for his kind hospitality during the course of this work.
\end{acknowledgments}

\end{document}